\documentclass[aps,10pt,pra,twocolumn,groupedaddress,floatfix]{revtex4-1}
\usepackage{graphicx}
\usepackage{subfigure}
\usepackage{amsfonts}
\usepackage{amsthm,amsbsy,amsmath,amsfonts,amssymb,amscd}
\usepackage{bm}
\usepackage{color}

\definecolor{burgundy}{rgb}{0.5, 0.0, 0.13}

\begin{document}


\title{Modeling Kleinian cosmology with electronic metamaterials}


\author{David Figueiredo, Felipe A. Gomes}
\affiliation{Departamento de F\'{\i}sica, CCEN,  Universidade Federal 
    da Para\'{\i}ba, Caixa Postal 5008,  58051-900 , Jo\~ao Pessoa, PB, Brazil }
\author{S\'ebastien Fumeron, Bertrand Berche}
\affiliation{ Statistical Physics Group, IJL, UMR Universit\'e de Lorraine - CNRS 7198
    BP 70239,  54506 Vand\oe uvre les Nancy, France}
\author{Fernando Moraes}
\affiliation{Departamento de F\'{\i}sica, CCEN, Universidade Federal da Para\'{\i}ba,
Caixa Postal 5008, 58051-900, Jo\~ao Pessoa, PB, Brazil }
\affiliation{Departamento de F\'{\i}sica, Universidade
    Federal Rural de Pernambuco, 
    52171-900 Recife, PE, Brazil}

\date{\today}

\begin{abstract}
This paper deals with the propagation of Klein-Gordon particles in flat background spacetime
exhibiting discontinuous metric changes from a Lorentzian signature $(-,+,+,+)$ to a Kleinian
signature $(-,+,+,-)$. A formal analogy with the propagation of electrons at a junction between an
anisotropic semiconductor and an electronic metamaterial is presented. From that analogy, we study
the dynamics of these particles falling onto planar boundary interfaces between these two families
of media and show a mirror-like behavior for the particle flux. Finally, the case of a double
junction of finite thickness is examined and the possibility of tunneling through it is discussed. A physical link between the metamaterial and the Kleinian slabs is found by calculating the time of flight of the respective traversing particles.
\end{abstract}



\maketitle

\section{Introduction}
Differential equations are a powerful way of describing most physical
phenomena. When the same equation fortunately describes very different systems, we have the rare
opportunity of building analogue models \cite{barcelo2005analogue}, sometimes allowing experiments to be made in a more
accessible system in order to understand better its less accessible counterpart. This
paper analyzes the propagation of plane waves in two very different systems united by the same
differential equation. They are, respectively, the propagation of relativistic spinless particles
(through an analysis of the Klein-Gordon equation, named after Oskar Klein and Walter Gordon) across
a Kleinian spacetime (named after Felix Klein) in an otherwise Lorentzian spacetime, and the
propagation of
ballistic electrons in a semiconductor across a region where their effective mass is
negative. 

The Klein-Gordon (KG) scalar field  has been used by several authors (see for example Refs. \cite
{horowitz1995quantum,malkiewicz2006probing,PhysRevD.87.063512}) as a theoretical probe for the
geometry of spacetime. This is important, for instance, in elucidating the role of singularities in
the dynamics of particles. Even though a classical test particle is directly affected by the
singularity, there are spacetimes  where a quantum particle feels, instead of the singularity, an
effective barrier \cite{horowitz1995quantum}. On the other hand, optical metamaterial modeling of
peculiar spacetimes like those of a black hole \cite{PhysRevD.82.124021,fernandez}, a spinning
cosmic string \cite{Mackay20102305},  a  multiverse \cite{2040-8986-13-2-024004}, or electromagnetic
black holes and magnetic monopoles \cite{PhysRevLett.99.183901} has become quite fashionable lately.
In particular, we note the study of   metric signature change \cite
{PhysRevLett.105.067402,PhysRevE.89.033202,Smolyaninov2013353} using optical metamaterials. The ease
of experimental verification of theses models is their greater asset, making cosmological
experiments a possibility in the optics laboratory. Electronic metamaterials  \cite
{dragoman2007metamaterials,Ding2013,PhysRevB.90.035138} are also an exciting possibility for modeling the of gravitational and cosmological problems. Contrasting with their optical counterparts, these metamaterials use ballistic electrons as their fundamental objects. In this article, to the best of our knowledge, we are the first to use an electronic metamaterial to model a cosmological problem: a spacetime with a discontinuous metric signature change. We show that the equation and boundary conditions governing the dynamics of a Klein-Gordon particle in such a spacetime are the same as those of a ballistic electron in an electronic metamaterial, when they move through a junction of different metric signature/electron effective mass. We use this result to study tunneling through a slab of Kleinian spacetime/negative electron effective mass.

In what follows we describe the two systems studied,
starting with the cosmological problem of a metric signature change from Lorentzian to Kleinian
along a chosen direction. Following this, we describe  its electronic metamaterial counterpart,
partially inspired by Ref. \cite{dragoman2007metamaterials}. Since we are
studying junctions of distinct spacetime structures and materials with different effective masses
for the electrons, it is necessary to establish appropriate boundary conditions. This is done in the
following section, which establishes the desired boundary conditions both for Klein-Gordon particles
and the electrons at the junction (in the latter, the mass depends on the position and therefore
has to be treated as an operator). Tunneling through a slab of Kleinian
signature is then considered next and, finally, we present our conclusions. 
\section{Klein-Gordon particles propagating in a spacetime with metric signature change} \label{sec:I}
A metric signature change from Euclidean $(+,+,+,+)$ to Lorentzian $(-,+,+,+)$ might have happened
in
the early Universe as a theoretical consequence of the ``no boundary proposal" of Hartle and Hawking
\cite{PhysRevD.28.2960,alty1994kleinian}. Another possibility, suggested by Sakharov
\cite{Sakharov:1984ir}, is that of a transition to a Kleinian $(-,+,+,-)$ signature. Metric
signature transitions also appear in the context of loop quantum gravity
\cite{1475-7516-2015-08-052} and a condensed matter experiment has been proposed to verify
``cosmological" particle production due to a signature change transition \cite{PhysRevD.76.124008}.
The effects of a Lorentzian to Kleinian transition on matter and electromagnetic waves was studied
in
detail by Alty \citep{alty1994kleinian} in 1994. We summarize below the case of a Klein-Gordon
particle crossing a Kleinian slab of spacetime.

Let us consider a signature changing spacetime described by the line element $ds^2=g_{\sigma\nu}dx^{\sigma}dx^{\nu}$:
\begin{equation}
    ds^2=\begin{cases}
        -c^2dt^2+dx^2+dy^2+dz^2 & \text{ for } z<0 \\ 
        -c^2dt^2+dx^2+dy^2-dz^2 & \text{ for } 0<z<l \\ 
        -c^2dt^2+dx^2+dy^2+dz^2 & \text{ for } z>l .
    \end{cases}
\label{sig_change}
\end{equation}
The dynamics of a relativistic spinless particle of mass $m$ is given by the Klein-Gordon equation
\begin{equation}
    \left(\Box - \mu^2\right)\phi =0, \label{KGeq}
\end{equation}
with $\Box = g^{\sigma \nu}\partial_{\sigma} \partial_{\nu}$ and $\mu=mc/\hbar$. Thus, in the 
Lorentzian regions
\begin{equation}
    \Box = -\frac{1}{c^2} \frac{\partial^2}{\partial t^2} + \frac{\partial^2}{\partial x^2} + \frac{\partial^2}{\partial y^2} + \frac{\partial^2}{\partial z^2}
\end{equation}
is the usual d'Alembertian operator, whereas in the Kleinian region
\begin{equation}
    \Box = -\frac{1}{c^2} \frac{\partial^2}{\partial t^2} + \frac{\partial^2}{\partial x^2} + \frac{\partial^2}{\partial y^2} - \frac{\partial^2}{\partial z^2}.
    \label{box}
\end{equation}

\begin{figure}[t]
    \centering
    \includegraphics[width=1.0\columnwidth]{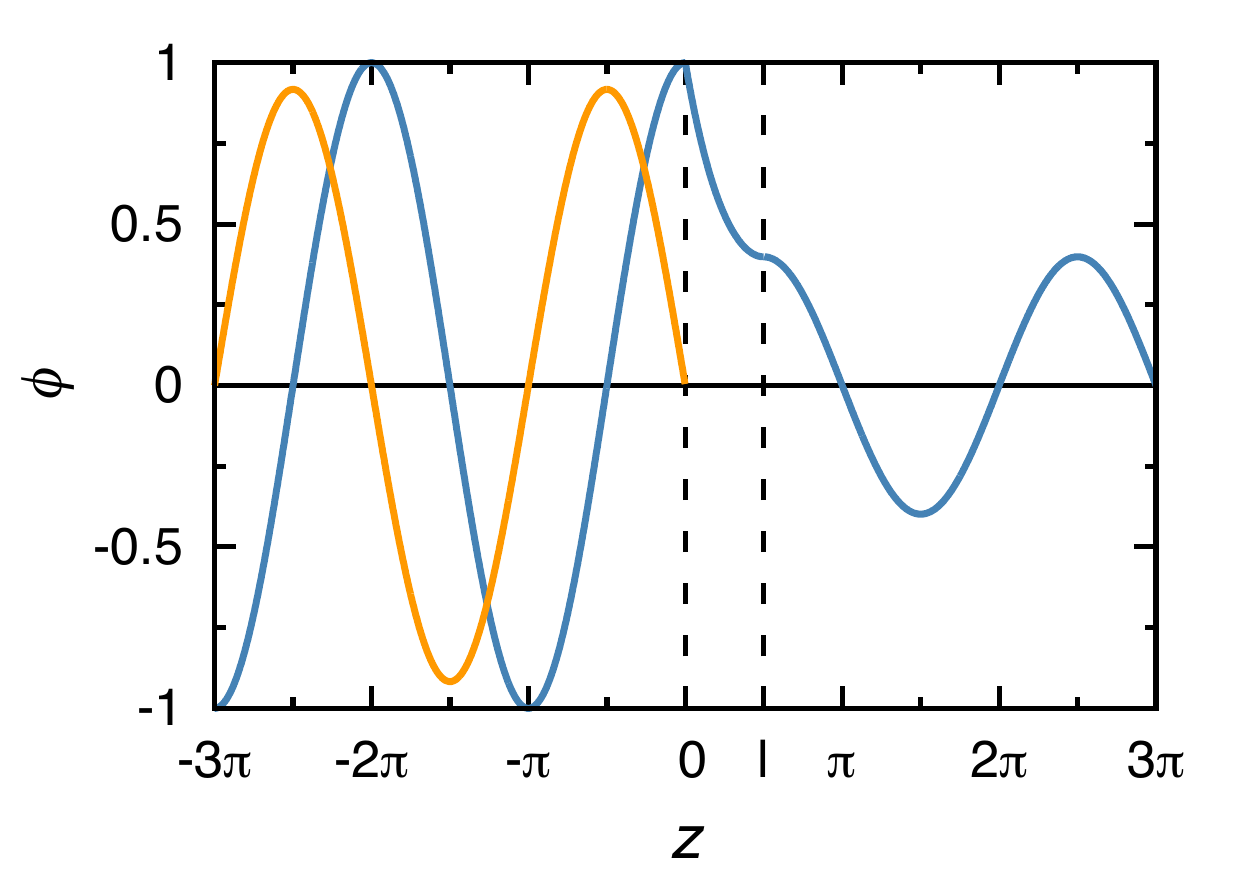}
    \includegraphics[width=1.0\columnwidth]{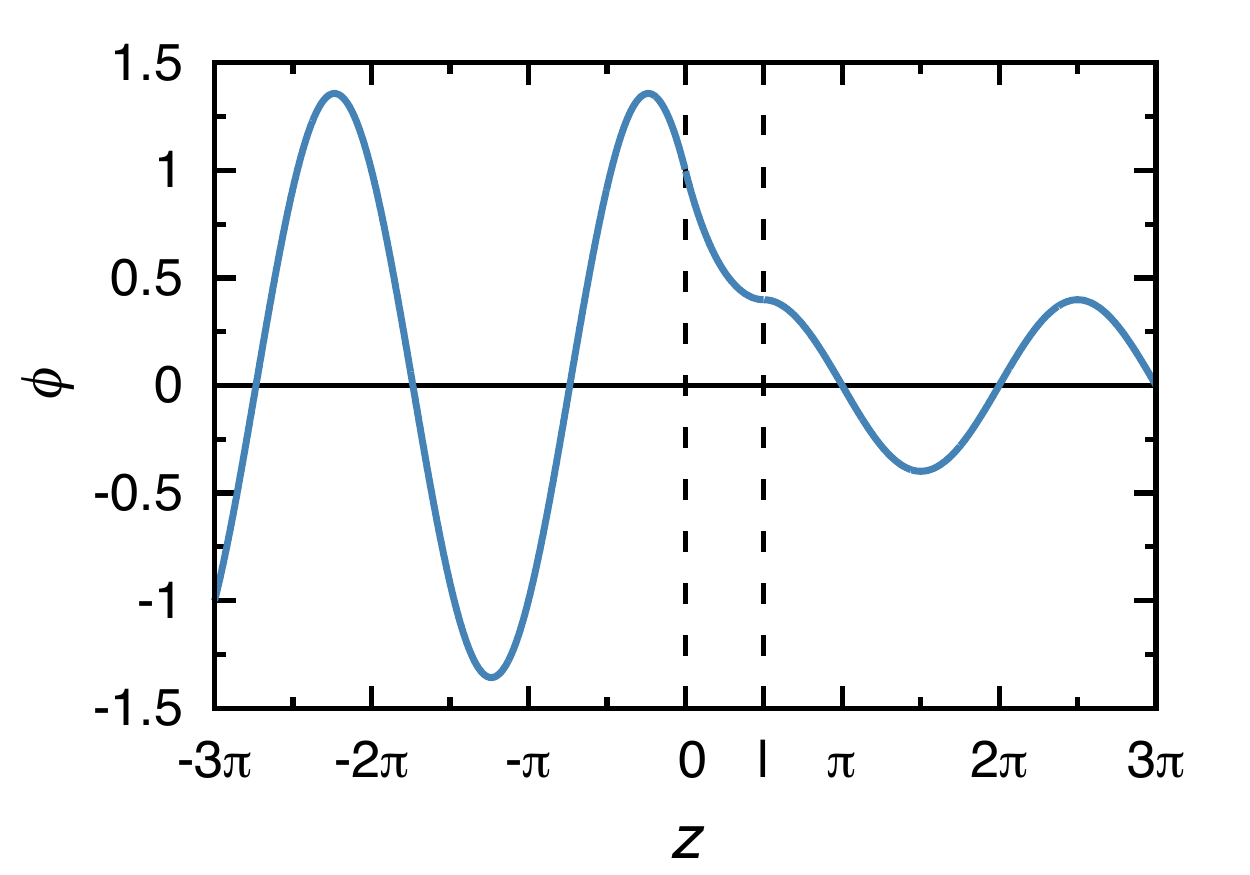}
    \caption{Top: Tunneling of the wave function $\phi$ (real part) through a Kleinian slab of
    length $l$: the blue line represents the incident and transmitted waves while the orange line
    denotes the reflected one. Bottom: Graph showing the total wave function for $z<0$ 
    (incident+reflected). The probability current is conserved since in Eq. (\ref{alty_coeff}) we
    have $\left|r\right|^2+\left| t\right|^2=1$. In both graphs, $z$ is given in units of $k_z^{-1}$.}
    \label{kleinian_fig}
\end{figure}

The solution of Eq. (\ref{KGeq}) in the Lorentzian regions is the plane wave
\begin{equation}
    \phi = e^{i \left(\mathbf{k}\cdot\mathbf{r}-\omega t \right)},
\end{equation}
where $\mathbf{r}=(x,y,z)$ and $\mathbf{k}=(k_x,k_y,k_z)$ is the wave vector. The angular frequency $\omega$ is given by the dispersion relation
\begin{equation}
    \frac{\omega^2}{c^2}= k_x^2 + k_y^2 +k_z^2 + \mu^2 . \label{dispL}
\end{equation}
Assuming a plane wave solution in the Kleinian region we see that the dispersion relation becomes
\begin{equation}
    \frac{\omega^2}{c^2}= k_x^2 + k_y^2 -k_z^2 + \mu^2  \label{dispK}
\end{equation}
meaning that $k_z$ must be purely imaginary in this region. In order to obtain the  dispersion
relation in the same form as Eq. $(\ref{dispL})$,  we made $k_z\rightarrow ip$ ($p\in \mathbb{R}$)
in the slab, such that
\begin{equation}
    \frac{\omega^2}{c^2}= k_x^2 + k_y^2 +p^2 + \mu^2 . \label{dispp}
\end{equation}
Therefore, the solution of the wave equation in the Kleinian region is
\begin{equation}
    \phi = e^{i(k_x x + k_ y y-\omega t) \pm p z}.
    \label{ansatz_klein}
\end{equation} 
Then, for the spacetime with signature transitions given by Eq. $(\ref{sig_change})$, we have
\begin{equation}
    \phi=e^{i(k_x x + k_ y y-\omega t)}\begin{cases}
        e^{ipz}+re^{-ipz}& \text{ for } z<0 \\ 
        ae^{pz}+be^{-pz}& \text{ for } 0<z<l \\ 
        te^{ipz}& \text{ for } z>l. 
    \end{cases}
    \label{scalar_field_sol}
\end{equation}
where $a$ and $b$ are complex amplitudes and  $r$ and $t$ are the reflected and transmitted
amplitudes from the slab. As obtained by Alty \cite{alty1994kleinian}, applying the usual boundary
conditions of continuity of both $\phi$ and $\partial \phi/\partial z$ at each interface, one gets
\begin{equation}
    \begin{aligned}
        r &=-i\text{tanh}(pl), \\ 
        a &= \left(\frac{1+i}{2}\right)\frac{e^{-pl}}{\text{cosh}(pl)},\\ 
        b &= \left(\frac{1-i}{2}\right)\frac{e^{pl}}{\text{cosh}(pl)},\\ 
        t &= \frac{e^{-ipl}}{\text{cosh}(pl)}.
    \end{aligned}
    \label{alty_coeff}
\end{equation}
Thus, there is tunneling across the Kleinian slab and it is easily verified that the probability
current is conserved by summing the squared modulus of the amplitude of the incident, reflected and
transmitted waves (see Fig. \ref{kleinian_fig}). 

However, looking more carefully at the spacetime structure given by Eq. (\ref
{sig_change}) we can see that the $g_{zz}$ metric component can be written as
\begin{equation}
    g_{zz}=H(-z)-H(z)+2H(z-l),
    \label{gzz_metric}
\end{equation}
where $H(z)$ is the Heaviside step function. It will be shown later (see Sec. \ref
{section_V}) that this discontinuous behavior of $g_{zz}$ leads to another condition for $\partial
\phi/\partial z$ at the junctions $z=0$ and $z=l$, giving us a different solution than Eq. (\ref
{alty_coeff}). For clarity, we will postpone the solution until Sec. \ref
{section_VI}.

\section{A ballistic electron metamaterial}
Let us consider a non-Euclidean geometry endowed with metric tensor $g_{ij}$. Therefore, the Schr\"
{o}dinger equation for electrons inside such a geometry is:
\begin{equation}
    -\frac{\hbar^2}{2m_e}\frac{1}{\sqrt{g}}\partial_i\left(\sqrt{g}g^{ij}\partial_j\right) \Psi+V\Psi=E\Psi, \label{ef-geom-schro}
\end{equation}
where $m_e=9,1\times10^{-31}$ kg is the true mass of an electron in vacuum and $g=\text{det}\:g_{ij}$.

The energy levels of spinless electrons in anisotropic semiconductors are governed by the effective mass Schr\"{o}dinger equation:
\begin{equation}
    -\frac{\hbar^2}{2}\left[\frac{1}{m^*}\right]^{ij}\partial_i\partial_j \Psi+V\Psi=E\Psi, \label{ef-mass-schro}
\end{equation}
where the effective mass tensor $[1/m^*]^{ij}$ is given by
\begin{equation}
\left[\frac{1}{m^*}\right]^{ij}=\left(\frac{1}{\hbar^2}\frac{\partial^2 E}{\partial k_i\partial k_j}\right)_{\mathbf{k}=0}=\begin{pmatrix}
    m_{11}^{-1}     & 0             & 0 \\
    0               & m_{22}^{-1}   & 0 \\
    0               & 0             & m_{33}^{-1}
\end{pmatrix}. 
\label{emass}
\end{equation}
The principal masses $m_{11}$, $m_{22}$ and $m_{33}$ can be of positive or negative sign \cite
{kittel2005introduction}. Comparing the kinetic terms between Eqs. (\ref{ef-geom-schro}) and (\ref
{ef-mass-schro}) shows that both equations coincide provided that $\partial_i \left(g^{ij}\sqrt
{g}\right)=0$. Then, the effective mass tensor generates an effective geometry, whose metric is 
\begin{equation}
    g^{ij}= m_e\left[\frac{1}{m^*}\right]^{ij}=\begin{pmatrix}
        \alpha_{11}     & 0               & 0 \\
        0               & \alpha_{22}     & 0 \\
        0               & 0               & \alpha_{33}
\end{pmatrix},
\label{mass_tensor}
\end{equation}
where $\alpha_{ii}=m_e/m_{ii}$.

For a model of an electronic metamaterial, we follow Ref. \cite{dragoman2007metamaterials},
where the effective mass $m^*$ and the difference $(E-V)$ are the electronic counterparts of the
electric permittivity $\epsilon$ and the magnetic permeability $\mu$, respectively. This analogy
is made in the sense that a positive (negative) effective mass corresponds to a positive (negative)
permittivity. The same is true for $(E-V)$ and $\mu$. As an
example, the electronic analogue of a hyperbolic metamaterial \cite{PhysRevA.92.063806} with permittivity tensor
\begin{equation}
    \epsilon_{ij}=\begin{pmatrix}
        \epsilon_1 & 0 & 0\\ 
        0 & \epsilon_1 & 0\\ 
        0 & 0 & -|\epsilon_2|
\end{pmatrix},
\label{epsilon_tensor}
\end{equation}
with $\epsilon_{xx}=\epsilon_{yy}=\epsilon_1>0,\epsilon_{zz}=\epsilon_2<0$, is the following
effective mass tensor:
\begin{equation}
    \left[\frac{1}{m^*}\right]^{ij}=\begin{pmatrix}
        m_1^{-1} & 0 & 0\\ 
        0 & m_1^{-1} & 0\\ 
        0 & 0 & -|m_2|^{-1}
\end{pmatrix},
\label{m_tensor}
\end{equation}
where $m_{11}=m_{22}=m_1>0,m_{33}=m_2<0$.

\begin{figure}[t]
    \centering
    \includegraphics[width=1.0\columnwidth]{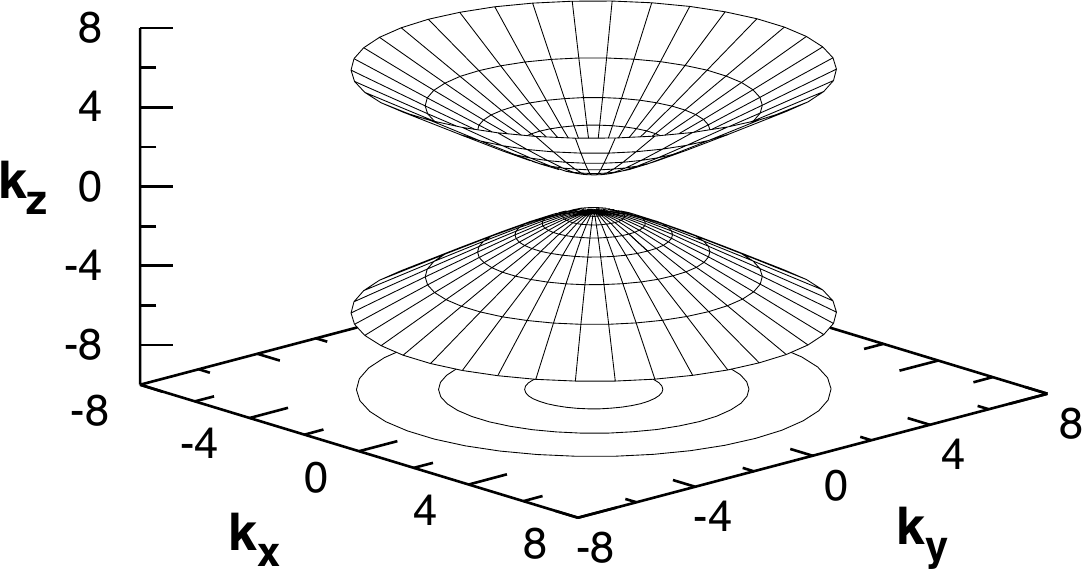}

    \vspace{0.5cm}

    \includegraphics[width=1.0\columnwidth]{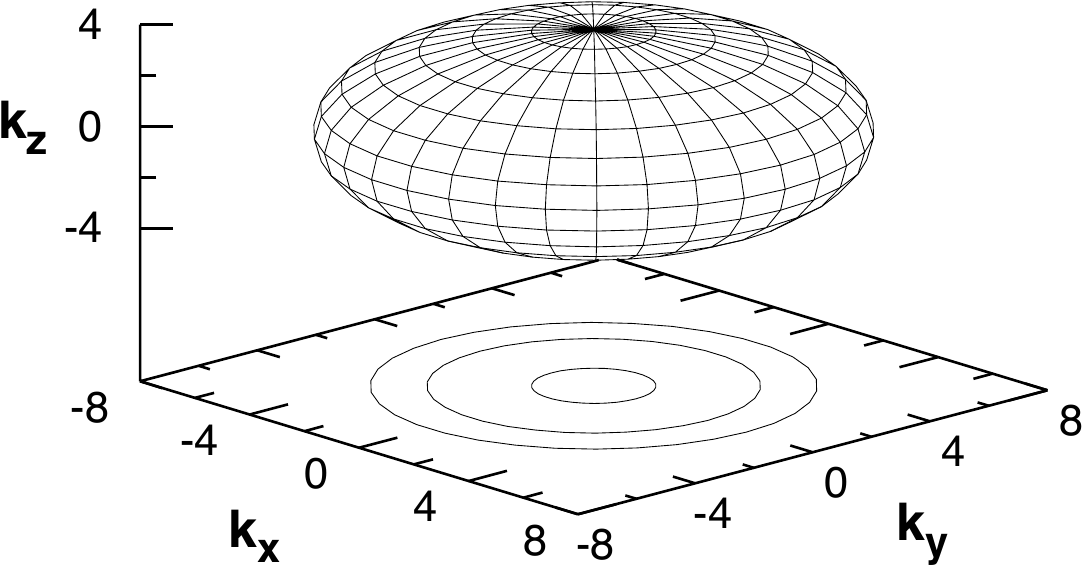}
    \caption{Top: A hyperboloid of two sheets for a constant frequency $\omega$ in
    Eq. (\ref {dispersion_hyper}). Dispersion relations like these are features of
    hyperbolic metamaterials. Bottom: Ellipsoid of revolution featuring the dispersion
    relation (\ref{dispL2}) for ballistic electrons.}
    \label{disp_rel_graph}
\end{figure}

However, one must be careful with this analogy concerning the dispersion relation. It is well
known \cite{shekhar2014hyperbolic} that the dispersion relation of a metamaterial with permittivity
tensor (\ref{epsilon_tensor}) is
\begin{equation}
    \frac{{k_z}^2}{\left(\epsilon_1\omega^2/c^2\right)}-\frac{{k_x}^2+{k_y}^2}{\left( |\epsilon_2|\omega^2/c^2\right)}=1,
    \label{dispersion_hyper}
\end{equation}
which is a hyperboloid of two sheets in $\mathbf{k}$ space (Fig. \ref{disp_rel_graph}). Regarding
the dispersion relation for the matter waves, we will consider the uniaxial mass tensor
(\ref{m_tensor}), which according to Eq. (\ref{mass_tensor}) is equivalent to a contravariant metric
tensor $g^ {ij}$ with signature $(+,+,-)$. For a free electron of energy $E>0$, Eq. (\ref
{ef-mass-schro}) leads to
\begin{equation}
-\frac{\hbar^2}{2}\left(\frac{1}{m_1}\frac{\partial^2 \Psi}{\partial x^2}+\frac{1}{m_1}\frac{\partial^2 \Psi}{\partial y^2}-\frac{1}{|m_2|}\frac{\partial^2 \Psi}{\partial z^2}\right)=E\Psi,
\label{sch-hyp} 
\end{equation}
for which, by the arguments discussed in the Appendix, the ansatz
\begin{equation}
\Psi(x,y,z)= e^{i(k_{x}x+k_{y}y)\pm k_{z}z},
\label{ansatz}
\end{equation}
is a solution provided the following dispersion relation holds:
\begin{equation}
\frac{{k_z}^2}{\left(2|m_2|\omega/\hbar \right )}+\frac{{k_x}^2+{k_y}^2}{\left(
2m_1\omega/\hbar\right)}=1,
\label{dispL2}
\end{equation}
where we used the fact that $E=\hbar\omega$. Equation (\ref{dispL2}) is an ellipsoid in
$\mathbf{k}$ space (Fig. \ref{disp_rel_graph}), in contrast with the hyperboloid given by Eq. (\ref
{dispersion_hyper}).

Comparing the time-dependent Schr\"{o}dinger equation solution
\begin{equation}
    \Psi(x,y,z,t)=e^{i(k_x x+k_y y-\omega t)\pm k_z z},
    \label{TDS_sol}
\end{equation}
with Eq. (\ref{ansatz_klein}), we see that an effective mass tensor like Eq. (\ref{m_tensor})
leads to the same solution for the Klein-Gordon equation in a Kleinian spacetime $(-,+,+,-)$.
Therefore, we can ``mimic'' a Klein-Gordon particle by choosing a suitable $[1/m^*]^{ij}$.

\section{Discontinuous signature transition\label{section3}}
Let us now consider a semiconductor which presents an abrupt transition from a positive to a
negative effective mass, i.e., a junction of positive and negative effective mass semiconductor
materials. Keeping in mind Eq. (\ref{mass_tensor}), this can be achieved if we choose a signature
transition $(+,+,+)$ to $(+,+,-)$ for the metric $g_{ij}$. For a transition at $z=z_0$
\begin{equation}
    g_{ij}=\begin{cases}
        \text{diag}(+,+,+) & \text{ for } z<z_0 \\ 
        \text{diag}(+,+,-)& \text{ for } z>z_0 .
    \end{cases}
    \label{single_junction}
\end{equation}

As already discussed, a metric with signature $(+,+,-)$ leads to a Schr\"{o}dinger equation with the
same solution as the one for the Klein-Gordon equation in a Kleinian spacetime. However, because the
wave function cannot be unbounded as $z\rightarrow \infty$, we must have for $z>z_0$ just the
evanescent wave of Eq. (\ref{TDS_sol}):
\begin{equation}
    \Psi(x,y,z,t)=e^{i(k_x x+k_y y-\omega t)-k_z z}.
    \label{eva_sol}
\end{equation}

For $z<z_0$, with a metric signature $(+,+,+)$, we choose $m_{33}=|m_2|$ (recall that $m_2<0$) for 
Eq. (\ref{m_tensor}).
Then, the Schr\"{o}dinger equation becomes
\begin{equation}
    -\frac{\hbar^2}{2}\left(\frac{1}{m_1}\frac{\partial^2 \Psi}{\partial x^2}+\frac{1}{m_1}\frac
    {\partial^2 \Psi}{\partial y^2}+\frac{1}{|m_2|}\frac{\partial^2 \Psi}{\partial
    z^2}\right)=E\Psi.
\end{equation}
Therefore,
\begin{equation}
    \Psi(x,y,z,t)=e^{i(k_x x+k_y y+k_z z-\omega t)},
    \label{planewave_sol}
\end{equation}
where, once again, we obtain the same dispersion relation (\ref{dispL2}) for the plane wave 
(\ref{planewave_sol}). In the remainder of this paper we will use the following notation:
\begin{align}
    \bm{x}&=(x,y),\\
    \bm{k}&=(k_x,k_y),\\
    p&=k_z.
\end{align}

At last, combining Eqs. (\ref{eva_sol}) and (\ref{planewave_sol}), we get
\begin{equation}
    \Psi=e^{i(\bm{k}\cdot\bm{x}-\omega t)}\begin{cases}
        ae^{ipz}+be^{-ipz}& \text{ for } z<z_0 \\ 
        ce^{-pz}& \text{ for } z>z_0 
    \end{cases}
\label{alty_sol}
\end{equation}
where $a$, $b$ and $c$ are complex amplitudes. In Ref. \cite
{alty1994kleinian}, one can see that Eq. (\ref{alty_sol}) has the same form for a Klein-Gordon
particle
through a signature change $(-,+,+,+)$ to $(-,+,+,-)$ at $z=z_0$. However, although it is tempting
to apply the usual boundary conditions at the interface (continuity of both
the wave function and its first derivative) care should be taken because of the mass change across
$z=z_0$.
Junction conditions for solutions of the Klein-Gordon equation are still a fairly hot debate \cite
{Dray93, Hayward94, Dray95}, but in our case, we will focus on a comparison with Alty's results and
examine them from the standpoint of the electronic metamaterial analogue. Thus, the aim of the
following section is to establish the appropriate boundary conditions for the single junction
(both for cosmological and electronic cases).

\section{Single junction and the mirror effect}
\label{section_V}
As already pointed in Secs. \ref{sec:I} and \ref{section3}, a discontinuous behavior in a
junction leads to another boundary condition for the derivative. Concerning a Klein-Gordon particle,
we start with the usual KG equation (\ref{KGeq}) with $\square$ replaced by the Laplace-Beltrami
operator
\begin{equation}
    \square=\frac{1}{\sqrt{|g|}}\partial_{\sigma}\left(\sqrt{|g|}g^{\sigma\nu}\partial_{\nu}\right),
    \label{lapbelt_op}
\end{equation}
where $g=\text{det}\:g_{\sigma\nu}$ and we will take $g^{\sigma\nu}$ as
\begin{equation}
    g^{\sigma\nu}=\begin{pmatrix}
-1 & 0 & 0 & 0 \\ 
 0 & 1 & 0 & 0 \\ 
 0 & 0 & 1 & 0 \\ 
 0 & 0 & 0 & f(z)
\end{pmatrix}.
\label{general_metric}
\end{equation}
Therefore, Eq. (\ref{lapbelt_op}) becomes
\begin{equation}
    \square=-\frac{1}{c^2}\frac{\partial^2}{\partial t^2}+\frac{\partial^2}{\partial x^2}+\frac
    {\partial^2}{\partial y^2}+\frac{1}{\sqrt{|g|}}\partial_z\left(\sqrt{|g|}f(z)\partial_z\right).
    \label{lapbelt_op2}
\end{equation}
Choosing the ansatz $\phi(\bm{x},z,t)=e^{i(\bm{k}\cdot\bm{x}-\omega t)}\varphi(z)$ for
the
KG equation, we
will get
\begin{equation}
    \frac{\omega^2}{c^2}-\bm{k}^2-\mu^2+\frac{1}{\varphi(z)}\frac{1}{\sqrt{|g|}}\partial_z\left(\sqrt
    {|g|}f(z)\partial_z\varphi(z)\right)=0.
\end{equation}
In order to keep the dispersion relation (\ref{dispp}), we must have
\begin{equation}
    \frac{1}{\sqrt{|g|}}\partial_z\left(\sqrt{|g|}f(z)\partial_z\varphi(z)\right)=-p^2\varphi(z).
    \label{KG_zpart}
\end{equation}
Taking $|g|=\pm 1/f(z)$ in Eq. (\ref{KG_zpart}) and after some calculations, one gets
\begin{equation}
    \left[-\frac{f'(z)}{2}\frac{d}{dz}+\frac{d}{dz}\left(f(z)\frac{d}{dz}\right)\right]\varphi(z)=-p^2\varphi
    (z),
    \label{KG_zpart2}
\end{equation}
where the prime symbol denotes $d/dz$.

Finally, let us take our case of interest and consider a signature change $(-,+,+,+)$ to $
(-,+,+,-)$
at $z=z_0$. Thus
\begin{align}
    f(z)&=H(-z+z_0)-H(z-z_0),\\
    f'(z)&=-2\delta(z-z_0),
\end{align}
where $\delta(z)$ is the Dirac delta function. Then, Eq. (\ref{KG_zpart2}) becomes
\begin{equation}
    \delta(z-z_0)\frac{d\varphi}{dz}+\frac{d}{dz}\left(f(z)\frac{d\varphi}{dz}\right)=-p^2\varphi(z).
    \label{KG_zpart4}
\end{equation}
Now, we integrate Eq. (\ref{KG_zpart4}) in the interval ($z_0-\epsilon,z_0+\epsilon$) around
$z=z_0$. We
consider $\epsilon$ to be very small. Therefore,
\begin{equation}
    \frac{\varphi'(z_0^+)+\varphi'(z_0^-)}{2}+\left[f(z)\frac{d\varphi}{dz}\right]^{z_0+\epsilon}_{z_0-\epsilon}=-p^2\int_
    {z_0-\epsilon}^{z_0+\epsilon}{\varphi dz},
    \label{KG_preint}
\end{equation}
where when integrating the first term on the left side of Eq. (\ref{KG_zpart4}) we have used the
fact
that
\begin{equation}
    \varphi'(z)\delta(z-z_0)=\frac{\varphi'(z_0^+)+\varphi'(z_0^-)}{2}\delta(z-z_0).
\end{equation}
In the limit $\epsilon \rightarrow 0$, Eq. (\ref{KG_preint}) becomes
\begin{equation}
    \frac{\varphi'(z_0^+)+\varphi'(z_0^-)}{2}+f(z_0^+)\varphi'(z_0^+)-f(z_0^-)\varphi'(z_0^-)=0.
    \label{almost_BD}
\end{equation}
Using $f(z_0^+)=-1$ and $f(z_0^-)=1$ in Eq. (\ref{almost_BD}), we obtain
\begin{equation}
    \varphi'(z_0^+)=-\varphi'(z_0^-).
    \label{KG_BD_derivative}
\end{equation}
Then, the boundary condition for $\partial \phi/\partial z$ at the junction $z=z_0$ will be
\begin{equation}
    \left.\frac{\partial \phi}{\partial z}\right|_{z=z_0^+}=-\left.\frac{\partial \phi}{\partial
    z}\right|_{z=z_0^-}
    \label{KG_BD_derivative2}
\end{equation}

For the electronic case, first it is important to mention that Eq. (\ref{ef-mass-schro}) is
valid only for the half-spaces $z<z_0$ and $z>z_0$ separately. As can be seen \cite
{pos_dep_mass}, one has to treat the effective mass tensor $[1/{m^*}]^{ij}$ as an operator, and
therefore care must be taken to ensure that the Hamiltonian is Hermitian and obeys Galilean
invariance. Following Ref. \cite{pos_dep_mass} we choose the following form for the kinetic term of
the Hamiltonian:
\begin{equation}
    \hat{H}_{\text{kin}}=\frac{1}{2}\hat{p}_{i}\left[\frac{1}{m^*}\right]^{ij}\hat{p}_{j},
\label{jonas_H}
\end{equation}
which satisfies both requisites.
Thus, the effective mass Schr\"{o}dinger equation becomes
\begin{equation}
    -\frac{\hbar^{2}}{2}\partial_i\left(\left[\frac{1}{m^*}\right]^{ij}\partial_j \Psi\right)+V\Psi=E\Psi,
\label{jonas_H2}
\end{equation}
which clearly reduces to equation (\ref{ef-mass-schro}) for a position-independent  $[1/{m^*}]^{ij}$.

In our case $[1/{m^*}]^{ij}$ is diagonal for the entire space. Thus,
as the transition (\ref{single_junction}) only affects $m_{33}$ at $z=z_0$, let us focus on the $z$
direction, with a
mass changing from $m_1$ to
$m_2$ at $z=z_0$. Therefore:
\begin{align}
    \frac{1}{m(z)}&=\frac{1}{m_1}H(-z+z_0)+\frac{1}{m_2}H(z-z_0),\label{mass_function}\\
    \frac{d}{dz}\left[\frac{1}{m(z)}\right]&=\left(\frac{1}{m_2}-\frac{1}{m_1}\right)\delta(z-z_0).
    \label
    {dmass_function}
\end{align}
For a free particle ($E>0$), we choose the ansatz $\Psi(\bm{x},z)=e^{i\bm{k}\cdot\bm
{x}}\psi(z)$. Then, Eq. (\ref{jonas_H2}) leads to
\begin{equation}
    \frac{d}{dz}\left( \frac{1}{m(z)}\frac{d\psi}{dz}\right)=-P^2\psi(z),
\label{jonas_Hz}
\end{equation}
where $P$ is a separation constant with dimensions of $k_z m^{-1}$. Since Eq. (\ref{jonas_Hz}) is
similar to
Eq. (\ref{KG_zpart4}), we integrate it in a small interval $(z_0-\epsilon,z_0+\epsilon)$ around the
junction $z=z_0$. Therefore,
\begin{equation} 
    \left[ \frac{1}{m(z)}\frac{d\psi}{dz}\right]^{z_0+\epsilon}_{z_0-\epsilon}=-P^2\int_{z_0-\epsilon}^{z_0+\epsilon}{\psi dz}.
    \label{definite_int}
\end{equation}
In the limit $\epsilon\rightarrow0$, we find
\begin{equation}
    \frac{1}{m_2}\psi'(z_0^+)=\frac{1}{m_1}\psi'(z_0^-).
    \label{BD_derivative}
\end{equation}

Thus, taking Eq. (\ref{BD_derivative}) with $m_1=|m_2|$ and
$m_2=-|m_2|$ one gets the following boundary condition for $\partial\Psi/\partial z$:
\begin{equation}
    \left.\frac{\partial \Psi}{\partial z} \right |_{z=z_0^+}=- \left.\frac{\partial \Psi}{\partial z} \right |_{z=z_0^-}\label{BD_2},
\end{equation}
which is the same as the condition (\ref{KG_BD_derivative2}) obtained for $\partial\phi/\partial z$.
Therefore, since we have the same boundary conditions for both cases, the analogy between $\phi$ and
$\Psi$ is fully achieved. We remark that the   same boundary condition is achieved by doing the self-adjoint extension of the Hamiltonian \cite{0264-9381-13-5-024,0264-9381-12-9-001}.

\begin{figure}[t]
        \centering
        \includegraphics[width=1.0\columnwidth]{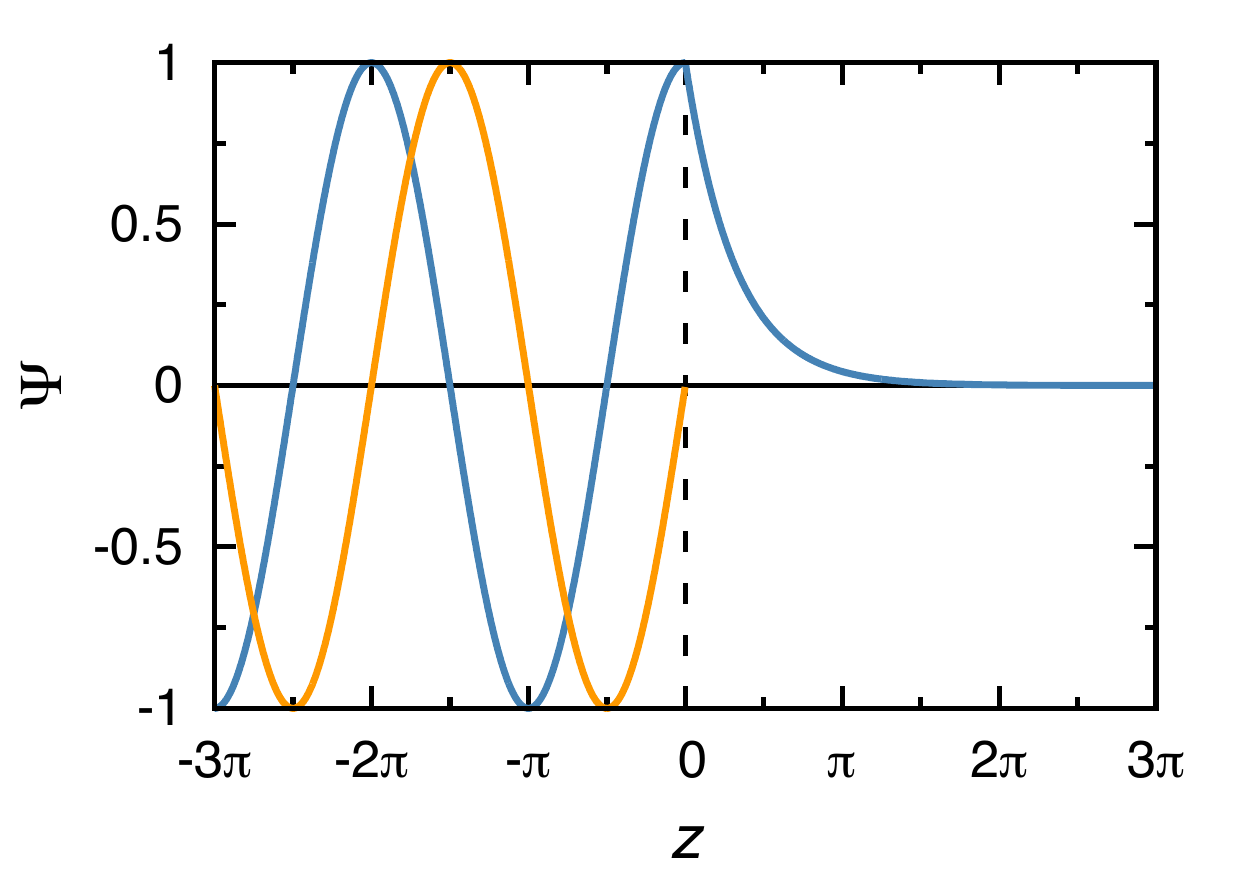}
        \caption{Behavior of the wave functions $\Psi$ and $\phi$ (real part)
        through an effective
        mass (signature) change at $z=0$. The incident wave (blue line) is completely reflected (orange line) with a $3\pi/2$ phase shift.  $z$ is given in units of $k_z^{-1}$.}
        \label{kleinian_fig2}
\end{figure}

Applying the continuity of the wave function and the boundary condition (\ref{BD_2}) for Eq. (\ref
{alty_sol}) we find $b=ia$ and $c=(1+i)a$. Then we compute the $j_z=(\hbar/m)\text{Im}
(\Psi^*\partial_z\Psi)$ component of the probability current for the incident, reflected and transmitted waves respectively:
\begin{align}
   j_0 &=\frac{\hbar\left | A \right |^2 }{|m_2|}p, \\ 
   j_r &=-\frac{\hbar\left | A \right |^2 }{|m_2|}p, \\ 
   j_t &= 0.
\end{align}
As one would expect for a transition from a Lorentzian signature $(-,+,+,+)$ to a Kleinian signature
$(-,+,+,-)$, the reflection coefficient $R=\left|j_r/j_0\right|=1$ and the incoming wave is totally
reflected with a phase shift of $3\pi/2$ (Fig. \ref{kleinian_fig2}) \cite{alty1994kleinian}. Such
an effect is analogous to optical reflection by semireflecting lossless mirrors \cite{Degiorgio}  and hence, the junction acts as a special  mirror for electronic probability currents.   
\section{A Wall of Kleinian Signature}
\label{section_VI}
Analogously to the frustration of evanescent waves in classical optics, the results of the previous sections (see also the Appendix) encourage us to investigate the tunneling through a Kleinian region. Therefore, consider a metamaterial (negative effective mass) slab with length $l$ sandwiched between two semiconductors with positive effective mass. The metric tensor $g^{ij}$ for this system is given by
\begin{equation}
    g^{ij}=\begin{cases}
        \text{diag}(+,+,+)& \text{ for } z<0 \\ 
        \text{diag}(+,+,-)& \text{ for } 0<z<l \\ 
        \text{diag}(+,+,+)& \text{ for } z>l 
    \end{cases}
\end{equation}

\begin{figure}[t]
    \centering
    \includegraphics[width=1.0\columnwidth]{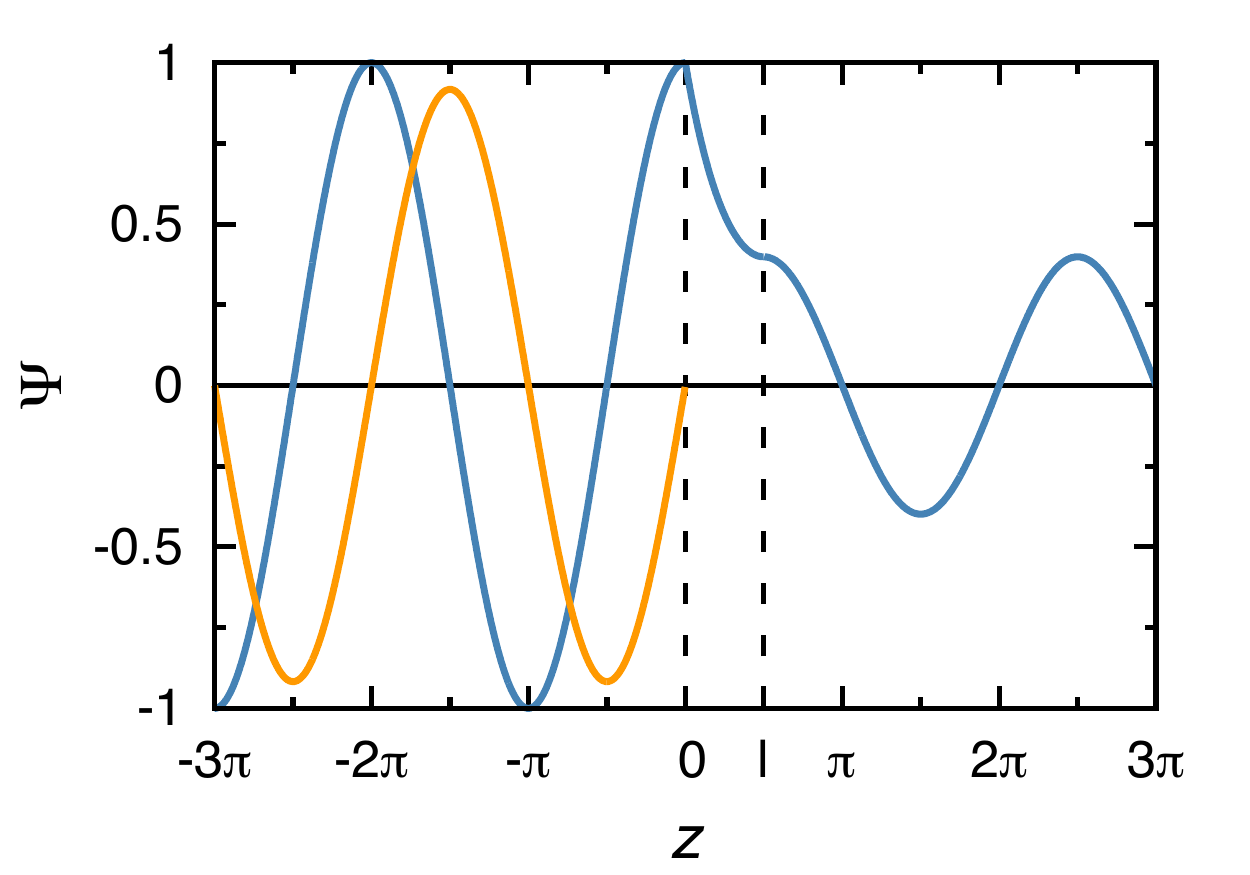}
    \includegraphics[width=1.0\columnwidth]{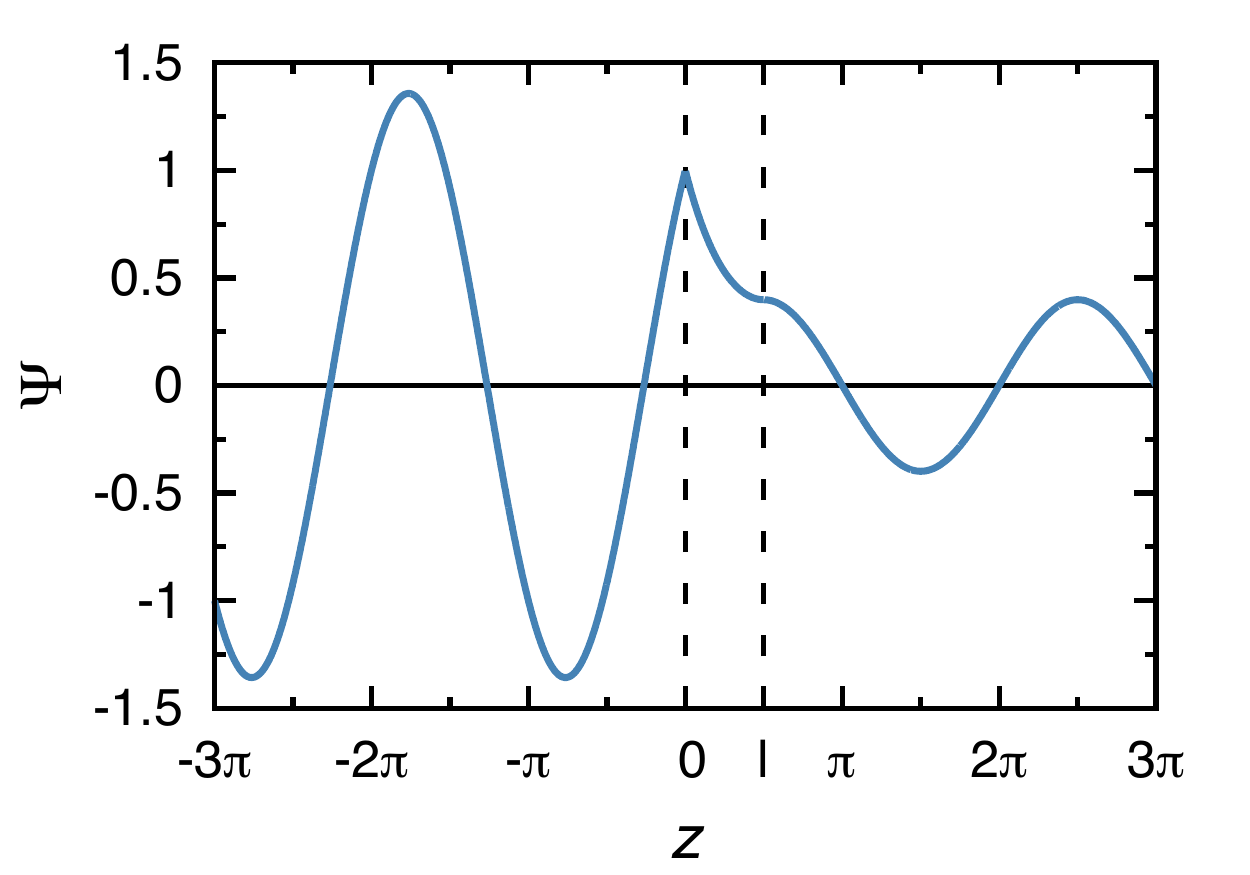}
    \caption{Top: Tunneling of the wave functions $\Psi$ and $\phi$ (real
    part) through a metamaterial (Kleinian spacetime) sandwich of length $l$:
    comparison
    with Fig. \ref
    {kleinian_fig} shows a phase shift of $\pi$ between the two reflected waves (orange lines). Bottom: Graph of the total wave function with a sharp peak at $z=0$. In both graphs, $z$ is given in units of $k_z^{-1}$.}
    \label{our_sandwich}
\end{figure}

\begin{figure}[t]
    \centering
    \includegraphics[width=1.0\columnwidth]{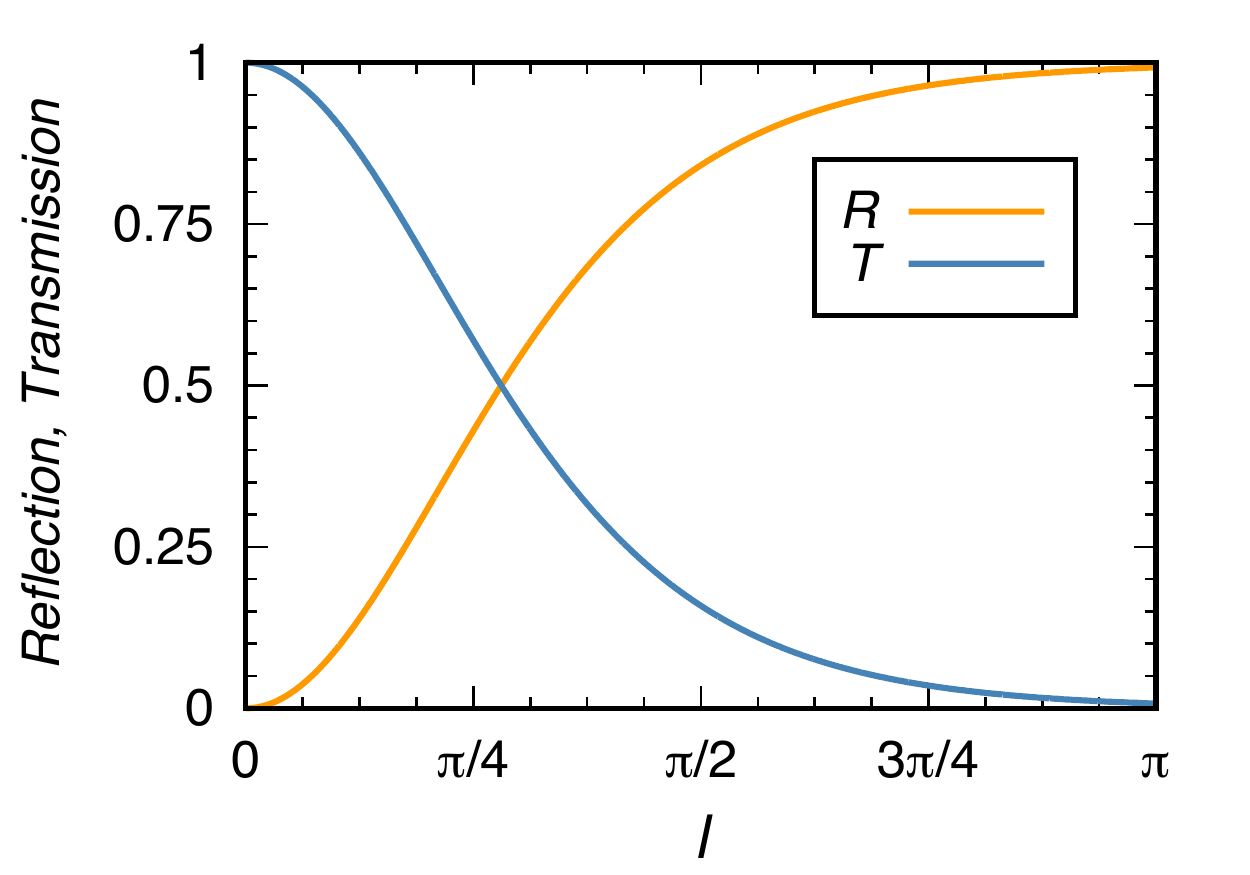}
     \caption{Graph showing the transmission and reflection coefficients as functions of the length $l$ of the slab (in units of $k_z^{-1}$). As we can see, the probability current is conserved besides the minor changes in the wave function $\Psi$.}
     \label{ref_vs_trans_graph}
\end{figure}

Following the same steps of Sec. \ref{section3}, we write the solution for $\Psi(\bm{x},z,t)$ as
\begin{equation}
    \Psi=e^{i(\bm{k}\cdot\bm{x}-\omega t)}\begin{cases}
        e^{ipz}+re^{-ipz}& \text{ for } z<0 \\ 
        ae^{pz}+be^{-pz}& \text{ for } 0<z<l \\ 
        te^{ipz}& \text{ for } z>l 
    \end{cases}
    \label{ballistic_sol}
\end{equation}
Applying the the continuity of $\Psi$ and Eq. (\ref{BD_2}) for $\partial \Psi/\partial z$ at $z=0$
and $z=l$, we find
\begin{equation}
\begin{aligned}
   r &=i\text{tanh}(pl), \\ 
   a &= \left(\frac{1-i}{2}\right)\frac{e^{-pl}}{\text{cosh}(pl)},\\ 
   b &= \left(\frac{1+i}{2}\right)\frac{e^{pl}}{\text{cosh}(pl)},\\ 
   t &= \frac{e^{-ipl}}{\text{cosh}(pl)}.
    \label{our_coeff}
\end{aligned}
\end{equation}
As we can see, except for slight changes, Eqs. (\ref{our_coeff}) are very similar to Eqs. 
(\ref{alty_coeff}). The differences arise because of the discontinuity of $\partial \Psi/\partial z$
in the boundary condition (\ref{BD_2}), which shifts the reflected wave by a phase of $\pi$ and
produces a ``sharp peak'' at $z=0$ in the wave function $\Psi$ (Fig. \ref{our_sandwich}).
However, besides those differences, the main physical properties remain the same as in Sec. \ref
{sec:I}. As an example, by calculating $j_0$, $j_r$ and $j_t$
\begin{align}
	j_0&=\frac{\hbar p}{|m_2|},\\
	j_r&=-\frac{\hbar p}{|m_2|}\text{tanh}^2(pl),\\
	j_t&=\frac{\hbar p}{|m_2|}\text{sech}^2(pl),\label{J_trans}
\end{align}
we can see explicitly that $j_0+j_r=j_t$ and the probability
current is still conserved. The reflection and  transmission   coefficients are then 
\begin{align}
	R&=\text{tanh}^2(pl),\\
	T&=\text{sech}^2(pl),
\end{align}
respectively. Their dependence on the slab length is shown in Fig. \ref{ref_vs_trans_graph}.

As remarked in Sec. \ref{sec:I}, the solution (\ref{alty_coeff}) for Klein-Gordon
particles through a slab of Kleinian spacetime has the same form as Eq. (\ref{ballistic_sol}).
Further,
by the similarity between Eqs. (\ref{KG_BD_derivative2}) and (\ref{BD_2}), the coefficients for $\phi$
must be those given by Eq. (\ref{our_coeff}). Thus, the graph for $\phi$ is the
same as that for $\Psi$ (see Fig. \ref{our_sandwich}).

Last, we obtain the time of flight  $\Delta t$ for a Klein-Gordon particle through a Kleinian
spacetime and the ballistic electrons through the metamaterial slab (length $l$). By definition
\begin{equation}
    \Delta t=\int_{0}^{l}{\frac{dz}{v_g(z)}},
    \label{trav_time_eq}
\end{equation}
where $v_g(z)=j_z/\rho$ is the group velocity and $\rho$ is the probability density. The expression
for $j_z$ is the same for both cases. However, the same is not true for $\rho$. Thus, for a Klein-Gordon
particle of mass $m$
\begin{equation}
    \rho=-\frac{\hbar}{mc^2}\text{Im}\left(\phi^{*}\partial_t \phi\right).
    \label{prob_dens_KG}
\end{equation}
Therefore, Eqs. (\ref{our_coeff}) and (\ref{J_trans}) give
\begin{equation}
    v_g(z)=\frac{pc^2}{\omega}\text{sech}\left[2p(z-l)\right].
    \label{group_velocity_KG}
\end{equation}
And after simple calculations, one gets
\begin{equation}
    \Delta t_{\text{KG}}=\frac{\omega}{2p^2c^2}\text{sinh}(2pl).
    \label{trav_time_KG}
\end{equation}

\begin{figure}[t]
    \centering
    \includegraphics[width=1.0\columnwidth]{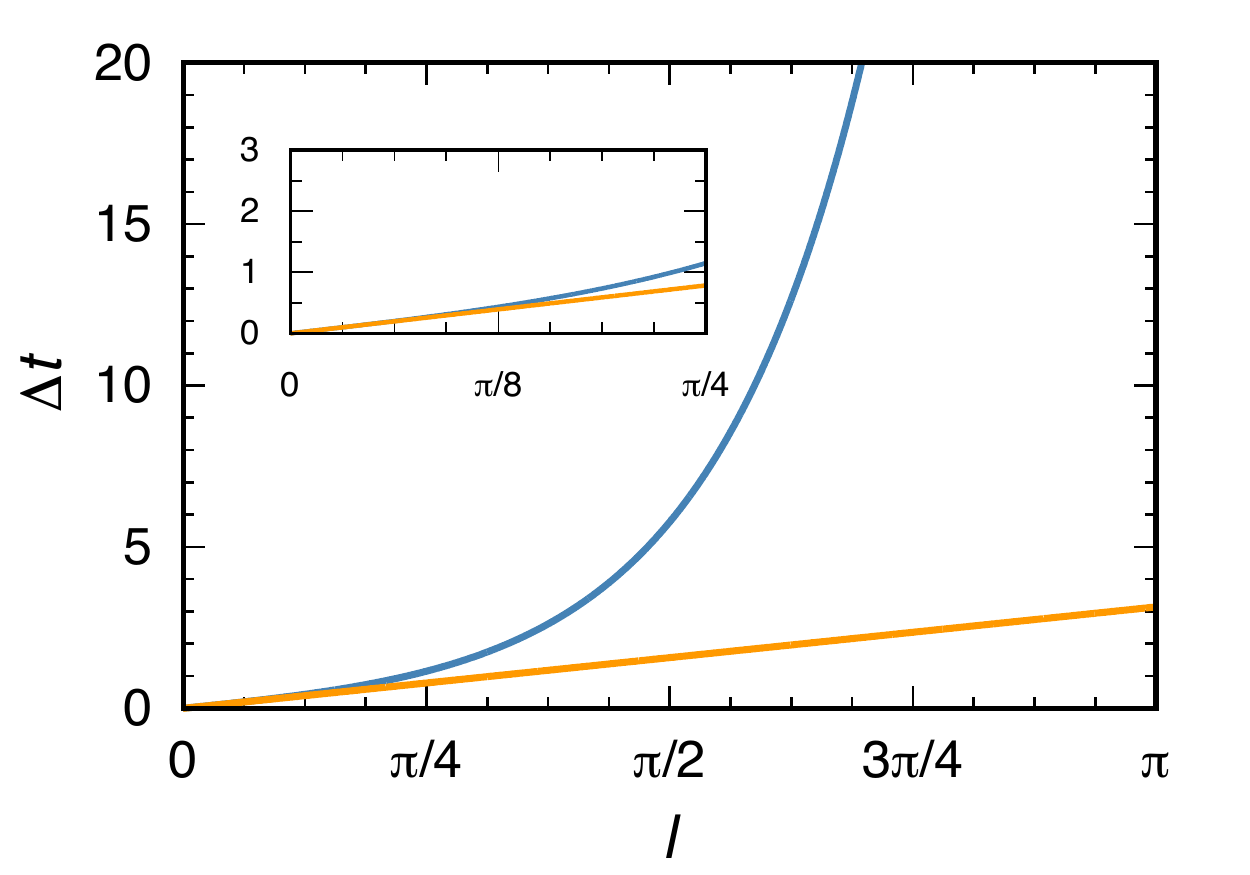}
    \caption{Time of flight of the ballistic electrons (KG particle) as a function of the slab
    length (Kleinian spacetime) $l$. The blue line shows the function given by Eqs. (\ref{trav_time_KG})
    and (\ref{trav_time_BE}). The orange one is for the case where the slab (Kleinian spacetime) is
    absent (where $v_g$ is given by $\hbar p/|m_2|$ for ballistic electrons and $pc^2/\omega$ for KG particles). In this graph, $l$ is given in units of $k_z^{-1}$ and $\Delta t$ in units of
    $|m_2|\hbar^{-1}k_z^{-2}$.}
     \label{trav_time_graph}
\end{figure}

Meanwhile, for ballistic electrons $\rho={|\Psi|}^2$. By the same procedure and
using Eqs. (\ref{our_coeff}) and (\ref{J_trans}), we will have for $v_g(z)$
\begin{equation}
    v_g(z)=\frac{\hbar p}{|m_2|}\text{sech}\left[2p(z-l)\right], \label{group_velocity_BE}
\end{equation}
and for the  time of flight
\begin{equation}
    \Delta t_{\text{BE}}=\frac{|m_2|}{2\hbar p^2}\text{sinh}(2pl).\label{trav_time_BE}
\end{equation}
The dependence of the time of flight on the slab length $l$ is shown in Fig. \ref{trav_time_graph}.
It is possible to establish a connection between $\Delta t_{\text{KG}}$ and
$\Delta t_{\text{BE}}$ through $\omega/p^2c^2\longleftrightarrow |m_2|/\hbar p^2$, or even $\hbar
\omega \longleftrightarrow |m_2|c^2$. This gives a numeric link between the two
systems, allowing experiments in an electronic metamaterial to be interpreted as analogous experiments with Klein-Gordon particles in Kleinian spacetime.

\section{Conclusions}
What is the behavior of a Klein-Gordon
particle crossing the boundary between a Lorentzian and a Kleinian spacetime? How long does it take
to tunnel through a slab of Kleinian spacetime? How could one perform experimental simulations of
such
particles? We answered these questions by investigating ballistic
electrons crossing the boundary between regions where they have masses of different signs. 
These quite different physical systems are described by the same solutions. Therefore, the second
case can be used to experimentally model  the first.  Following Dragoman and Dragoman \cite
{dragoman2007metamaterials}, we studied  the tunneling of ballistic electrons through a slab of
electronic metamaterial. As a further step, we improved the electronic metamaterial model by noticing that, since the effective mass depends on the position, it should be treated as an operator. Special boundary conditions at the interface \cite{pos_dep_mass} were then applied to both the single boundary and slab cases. The propagating wave corresponding to either case was then obtained, in perfect agreement with the expected behavior for Klein-Gordon particles in Kleinian spacetime. With this simple example we showed the power of electronic metamaterials to simulate cosmological situations, as already pointed out by Smolyaninov and Narimanov \cite{PhysRevLett.105.067402} for the electromagnetic ones. In the present case, in order to model Klein-Gordon particles in Kleinian spacetime with ballistic electrons in an electronic metamaterial, we need to choose a material with negative effective electronic mass and no bias voltage applied (or, if any, such that $E-V>0$) or one with positive effective mass and bias voltage higher than the kinetic energy of the electrons.
\begin{acknowledgments}
F.M. is thankful for the financial support and warm hospitality  of the group at Universit\'e de Lorraine where this work was conceived and partly done. This work has been partially supported by CNPq, CAPES and FACEPE (Brazilian agencies).
\end{acknowledgments}

\appendix*
\section{The one-dimensional Schr\"{o}dinger equation and the effective mass}
\label{appendix}
Here we summarize some simple but important concepts that are necessary for a better understanding
of this
paper. We start by looking at the Klein-Gordon equation in the Kleinian spacetime by rewriting Eq. 
(\ref
{KGeq}) with the d'Alembertian operator given by Eq. (\ref{box}). With a few manipulations and after
inserting the ansatz $\phi(x,y,z,t) = e^{-i\omega t}e^{i(k_x x + k_y y)}\varphi (z)$, one
easily gets
\begin{equation}
\frac{d^2 \varphi}{dz^2} = -\left(k_x^2 + k_y^2 +\mu^2 - \frac{\omega^2}{c^2}\right)\varphi = p^2 \varphi, \label{1DeqK}
\end{equation}
where the dispersion relation (\ref{dispp}) was used.

Now, let us consider the one-dimensional time-independent Schr\"{o}dinger equation for a
particle with effective mass $m^*$ moving along the $z$ axis, subjected to a constant potential $V$:
\begin{equation}
 -\frac{\hbar^2}{2m^*}\frac{d^2\Psi}{dz^2}+V\Psi=E\Psi, 
\end{equation}
or
\begin{equation}
    \frac{d^2\Psi}{dz^2}=-k^2\Psi,\label{1DeqM}
\end{equation}
where the constant $k$ is defined as 
\begin{equation}
    k=\frac{\sqrt{2m^*(E-V)}}{\hbar}.
    \label{k_constant}
\end{equation}
This has a  ``plane'' wave solution given by
\begin{equation}
    \Psi(z)=e^{\pm ikz}.
    \label{pseudo_plane}
\end{equation}
Since $m^*$ can be positive or negative and considering the respective signs of $m^*$ and $(E-V)$,
it is easily
verified that, if both quantities have the same sign, the wave number
$k$ in Eq. (\ref{k_constant}) is real and the solution (\ref{pseudo_plane}) is a plane wave.
However, if the signs are different, $k$ is purely imaginary, Eqs. (\ref{1DeqK}) and (\ref{1DeqM})
are identical, and Eq. (\ref{pseudo_plane}) is a real
exponential. Thus, making $k\rightarrow ip$ ($p\in \mathbb{R}$) in Eq. (\ref{pseudo_plane}) we
obtain
\begin{equation}
    \Psi(z)=e^{\pm pz},
    \label{real_sol}
\end{equation}
where $p$ is given by
\begin{equation}
    p=\frac{\sqrt{2\left|m^*(E-V)\right|}}{\hbar}.
\end{equation}

\bibliography{references}
\end{document}